%% file: npsp.tex
\documentclass[%
reprint, amsmath, amssymb, aps, ]{revtex4-1}

\usepackage{graphicx}
\usepackage{dcolumn}
\usepackage{bm}
\include{def-com}

\begin{document}
\preprint{APS/123-QED}

\title{\boldmath Determination of the number of $\psp$ events at BESIII }
\author{
\bcl
 M.~Ablikim$^{1}$, M.~N.~Achasov$^{5}$, O.~Albayrak$^{3}$, D.~J.~Ambrose$^{39}$, F.~F.~An$^{1}$, Q.~An$^{40}$, J.~Z.~Bai$^{1}$, Y.~Ban$^{27}$, J.~Becker$^{2}$, J.~V.~Bennett$^{17}$
% ...
\\
\vspace{0.2cm}
(BESIII Collaboration)\\
\vspace{0.2cm} {\it
$^{1}$ Institute of High Energy Physics, Beijing 100049, P. R. China\\
$^{2}$ Bochum Ruhr-University, 44780 Bochum, Germany\\
$^{3}$ Carnegie Mellon University, Pittsburgh, PA 15213, USA\\
% ...
$^{5}$ G.I. Budker Institute of Nuclear Physics SB RAS (BINP), Novosibirsk 630090, Russia\\
% ...
$^{17}$ Indiana University, Bloomington, Indiana 47405, USA\\
% ...
$^{27}$ Peking University, Beijing 100871, P. R. China\\
% ...
$^{39}$ University of Rochester, Rochester, New York 14627, USA\\
$^{40}$ University of Science and Technology of China, Hefei 230026, P. R. China\\
% ...
}\end{center}}

\begin{abstract}
The number of $\psp$ events accumulated by the BESIII experiment from
 March 3 through April 14, 2009, is determined by counting inclusive
 hadronic events. The result is $106.41\times(1.00\pm 0.81\%)\times
 10^{6} $. The error is systematic only; the statistical error is
 negligible.
\end{abstract}

 \maketitle
\section{Introduction}
In 2009, the world's largest $\psp$ sample to date was collected at
BESIII, allowing more extensive and precise studies of $\psp$
decays. The number of $\psp$ events, $N_{\psp}$, is important in all
$\psp$ analyses, including studies both of the direct decays of the
$\psp$, as well as its daughters, $\chicJ$, $h_c$, and $\eta_c$. The
precision of $\npsp$ will directly affect the precision of all these
measurements.

In this paper, we determine $\npsp$ with $\psp\ar inclusive~hadrons$,
whose branching ratio is known rather precisely, $(97.85\pm
0.13)$\%~\cite{PDG}. Also, a large off-resonance continuum data sample
at $E_{cm}=3.650$ GeV with an integrated luminosity of 44 pb$^{-1}$
was collected. These events are very similar to the continuum
background under the $\psp$ peak. Since the energy difference is very
small, we can use the off-resonance data to estimate this background.

BEPCII is a double-ring $e^+e^-$ collider designed to provide
$e^+e^-$ interactions with a peak luminosity of $10^{33}
~\rm{cm}^{-2}\rm{s}^{-1}$ at a beam current of 0.93 A. The
cylindrical core of the BESIII detector consists of a helium-based
main drift chamber (MDC), a plastic scintillator time-of-flight
system (TOF), and a CsI(Tl) electromagnetic calorimeter (EMC),
which are all enclosed in a superconducting solenoidal magnet
providing a 1.0 T magnetic field. The solenoid is supported by an
octagonal flux-return yoke with resistive plate counter muon
identifier modules interleaved with steel. The acceptance for
charged particles and photons is 93\% over 4$\pi$ stereo angle,
and the charged-particle momentum and photon energy resolutions at
1 GeV are 0.5\% and 2.5\%, respectively.

The BESIII detector is modeled with a Monte Carlo (MC) simulation
based on \textsc{geant}{\footnotesize 4}~\cite{geant4, geant42}.  For
the simulation of inclusive $\psp$ decays, we use the {\sc evtgen}
generator~\cite{evtgen}. Known $\psp$ decay channels are generated
according to branching ratios in the PDG~\cite{PDG}; the remaining
unknown decays are generated by the {\sc lundcharm} model~\cite{lund}.

\section{Event selection}
There are many types of events in the data collected at the $\psp$
energy point, including $\psp\ar$ hadrons and lepton pairs ($\EE,
~\MM$, and $\tau^+\tau^-$), radiative returns to the $\jpsi$, and
$\jpsi$ decays from the extended tail of the $\jpsi$ Breit-Wigner
distribution. In addition, there are non-resonance (QED)
processes, which make up the continuum background, including
$\EE\ar \gamma^* \ar$ hadrons, lepton pairs, and $\EE\ar\EE$ +X
(X=hadrons, ~lepton pairs). Non-collision events include cosmic
rays, beam-associated background, and electronic noise. The signal
channel is the process $\psp\ar$ hadrons.  The data collected at
the off-resonance energy include all of the above except $\psp\ar$
hadrons and lepton pairs.

Event selection includes track level selection and event
level selection. At the track level, good charged tracks are required
to pass within 1 cm of the beam line in the plane perpendicular to
the beam and within $\pm$15 cm from the Interaction Point (IP) in
the beam direction. Photon candidate showers reconstructed from
the EMC barrel region ($|\cos\theta|<0.8$) must have a minimum
energy of 25 MeV, while those in the end-caps
($0.86<|\cos\theta|<0.92$) must have at least 50 MeV. The showers
in the angular range between the barrel and end-cap are poorly
reconstructed and excluded from the analysis. Requirements on the
EMC cluster timing are applied to suppress electronic noise and
energy deposits unrelated to the event.

At the event level, at least one good charged track is required. If
the number of good charged tracks is larger than 2, i.e.
$N_{good}>2$, no additional selection is needed. If $N_{good}=2$,
where the Bhabha and dimuon events are dominant backgrounds, the
momentum of each track is required to be less than 1.7 GeV/$c$, and the
opening angle between the two tracks is required to be less than
$176^{\circ}$ to suppress these backgrounds. Figures~\ref{pvspbb} and
~\ref{pvspmc} show scatter plots of the momentum of one track versus
the momentum of the second track for MC simulated Bhabha events and
inclusive MC events with two charged tracks,
respectively. Figures~\ref{angbb} and ~\ref{angmc} show the opening
angle distributions of MC simulated Bhabha events and inclusive MC
events with two charged tracks, respectively. In addition,
$E_{visible}/E_{cm}>0.4$ is required to suppress low energy background
(LEB), comprised mostly of $\EE\ar\EE +X$ and double ISR events ($\EE
\to \gamma_{ISR} \gamma_{ISR} X$). Here, $E_{visible}$ denotes the
visible energy which is defined as the energy sum of all charged
tracks (calculated with the track momentum and assuming a $\pi^{\pm}$
mass) and neutral showers, and $E_{cm}$ denotes the center-of-mass
energy. Figure~\ref{evis2prdt} shows the $E_{visible}$ distribution for
data and inclusive MC events with two charged tracks. The excess in data
at low energy is from the LEB.

If $N_{good}=1$, at least two additional photons are required in
an event. From all photon pair combinations, the combination whose
invariant mass, $M_{\GG}$, is closest to the $\pi^0$ mass is
selected, and $|M_{\GG}-M_{\pi^0}|<0.015$ GeV/$c^2$ is required.
$E_{visible}/E_{cm}>0.4$ is also required to suppress the LEB.
Figure~\ref{pi0} shows the $M_{\GG}$ distributions in the $\pi^0$
mass region for data and MC simulation. Figure~\ref{evis1prdt} shows the
$E_{visible}$ distribution for data and inclusive MC events. The
excess in data at low energy is from LEB.

\bfg
\includegraphics[width=5.0cm,height=5.0cm]{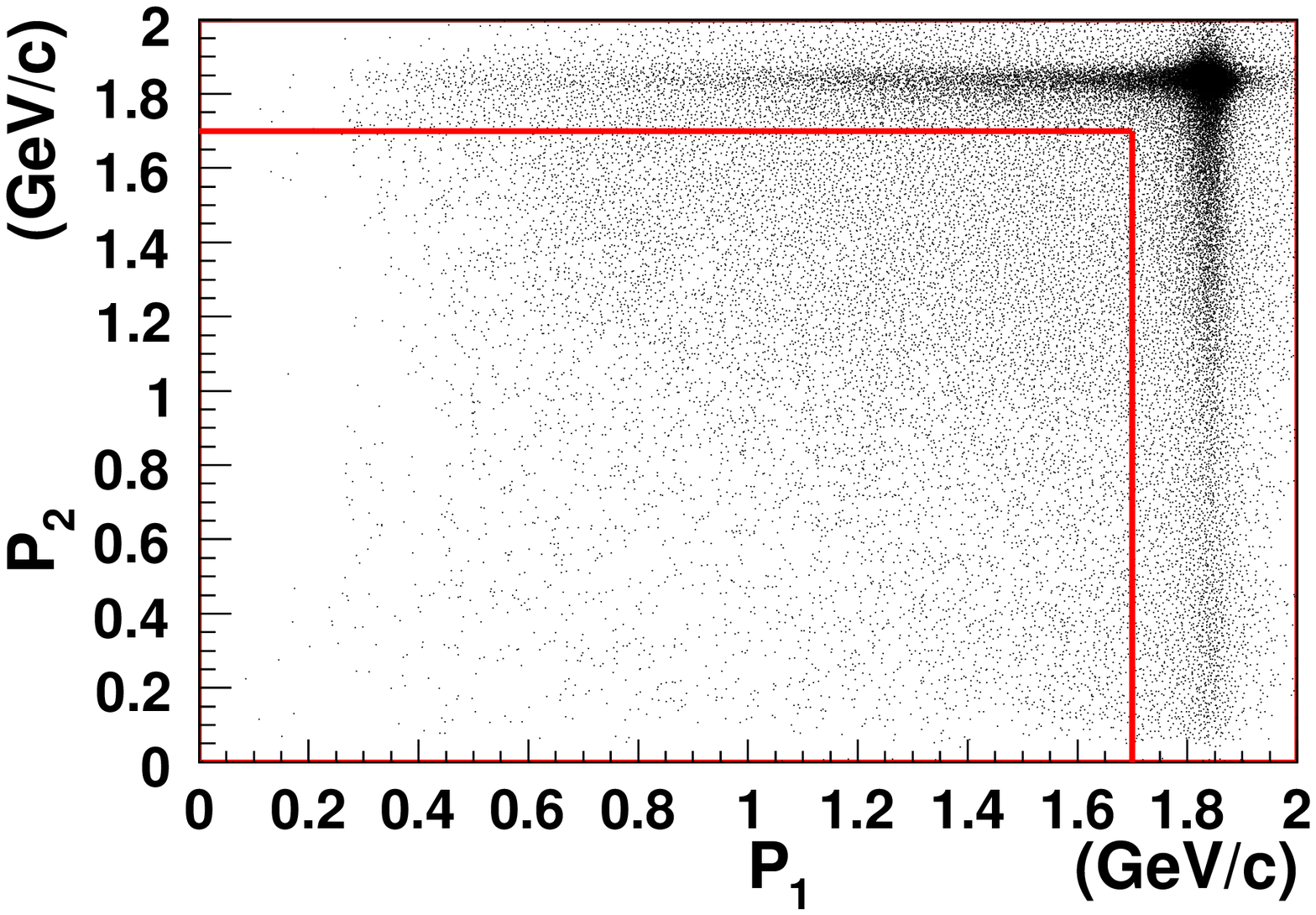}
\caption{\label{pvspbb}The distribution of $P_2$ versus $P_1$ from
MC simulated Bhabha events.  The horizontal and vertical lines show
the selection requirements to remove Bhabha and $e^+ e^- \to \mu^+
\mu^-$ events.}
\efg

\bfg
\includegraphics[width=5.0cm,height=5.0cm]{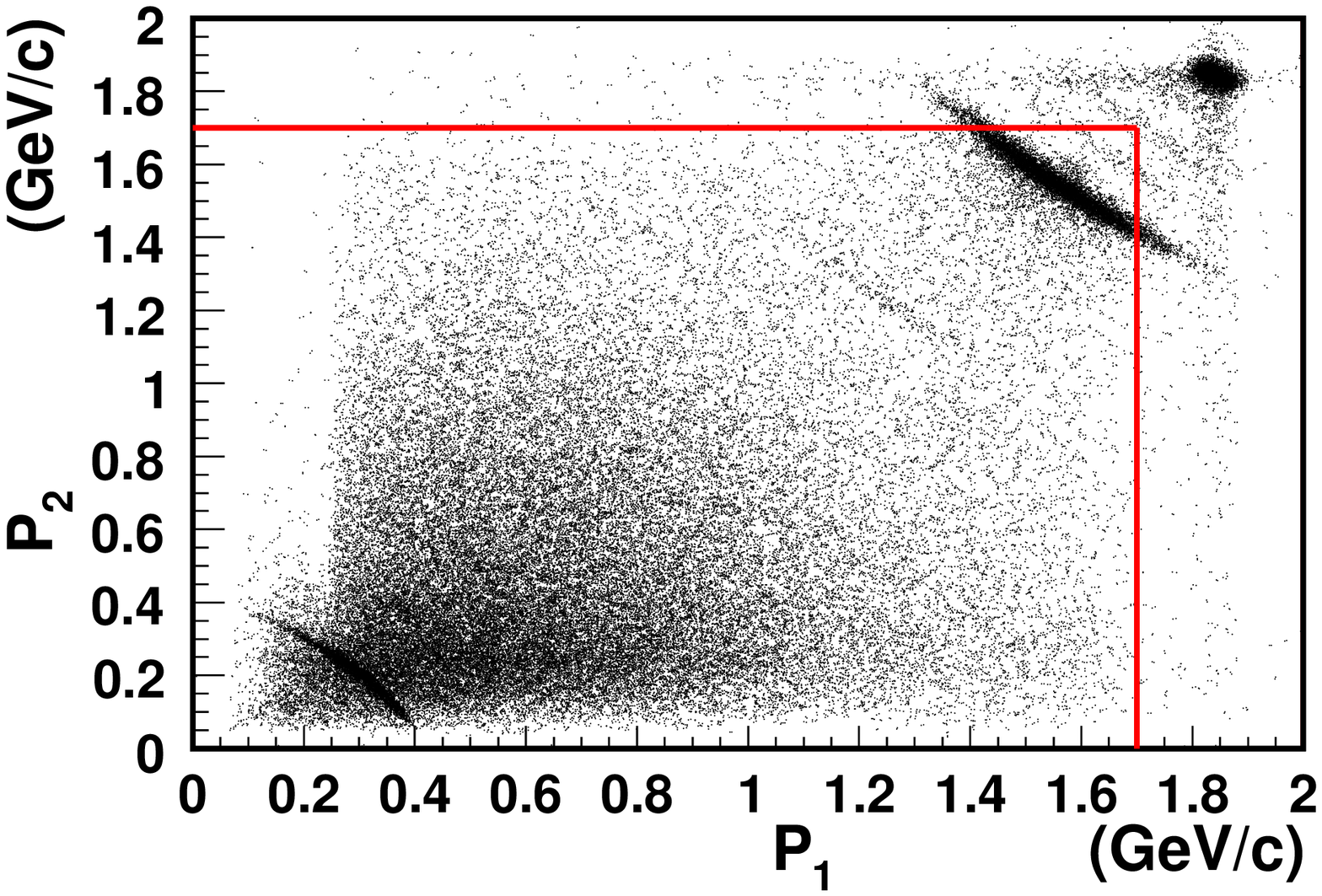}
\caption{\label{pvspmc}The distribution of $P_2$ versus $P_1$
from inclusive MC events with two charged tracks. The horizontal
and vertical lines show the selection requirements to remove
Bhabha and $e^+ e^- \to \mu^+ \mu^-$ events.}  \efg

\bfg \includegraphics[width=5.0cm,height=4.5cm]{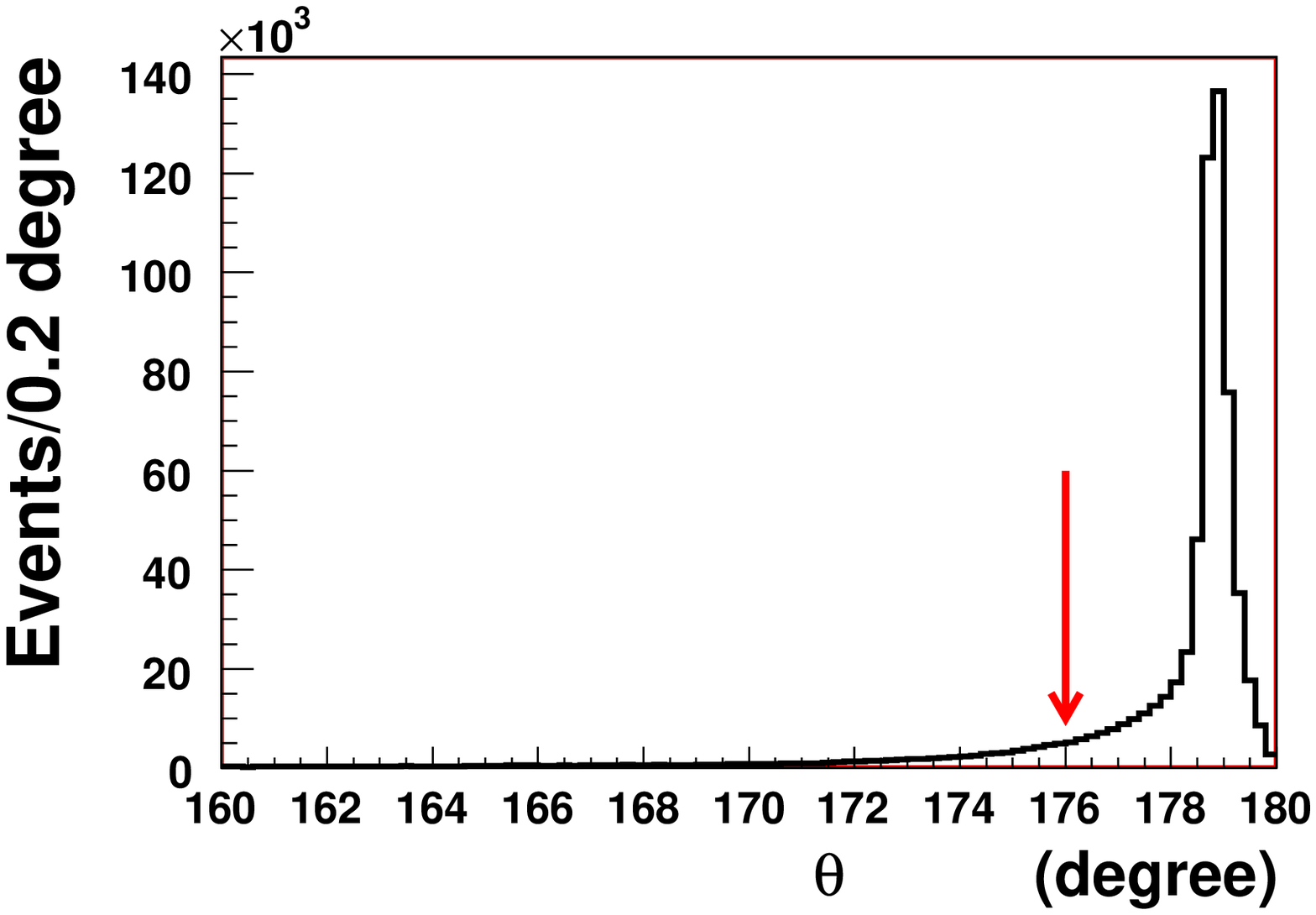}
\caption{\label{angbb}The distribution of angle between tracks
for MC simulated Bhabha events. The arrow shows the angle
requirement used to remove most Bhabha events.}  \efg

\bfg \includegraphics[width=5.0cm,height=4.5cm]{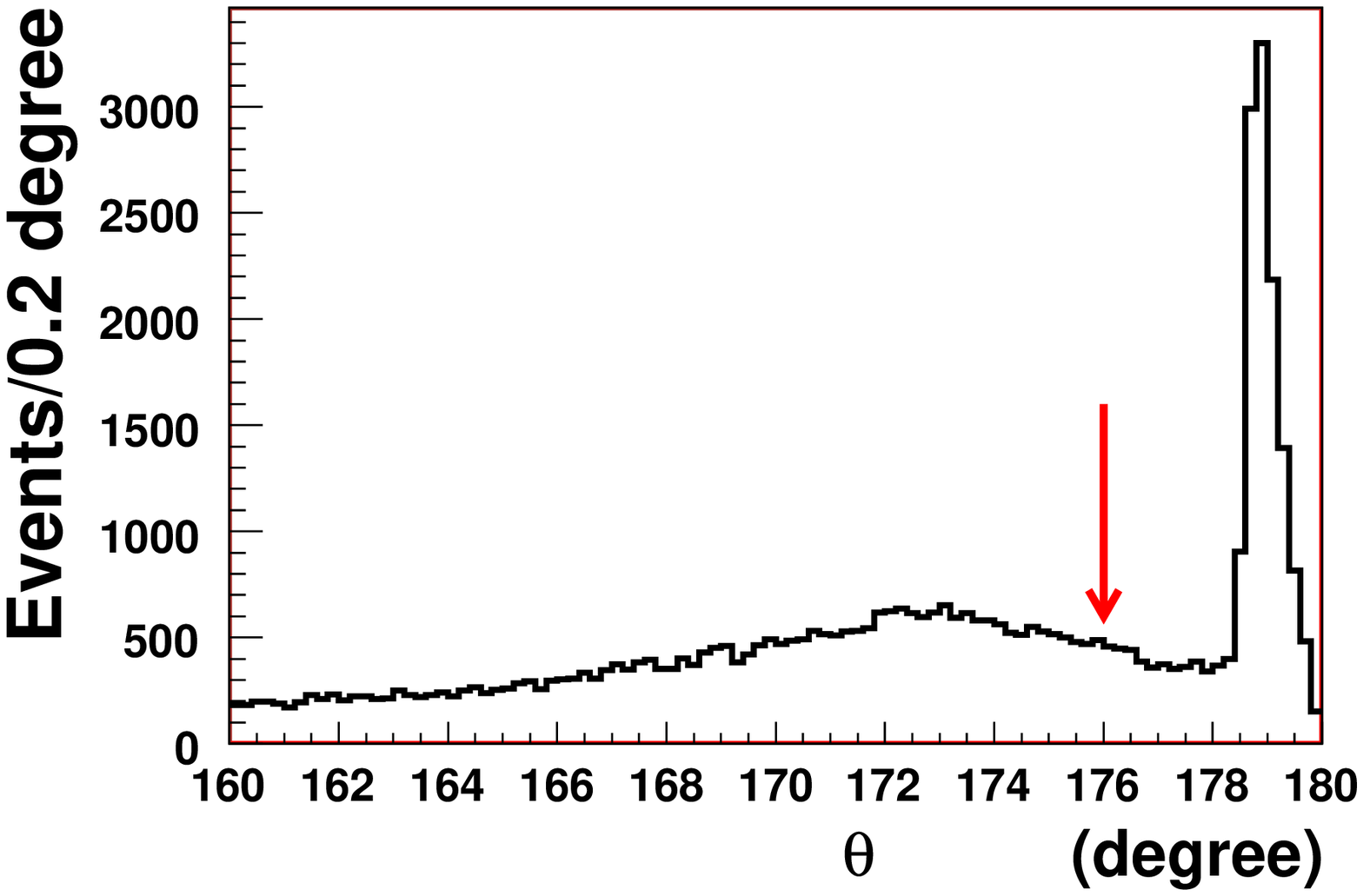}
\caption{\label{angmc}The distribution of angle between tracks
for inclusive MC events with two charged tracks.  The arrow shows
the angle requirement used to remove most Bhabha events.}  \efg

\vspace{0.5 in} \bfg
\includegraphics[width=5.0cm,height=4.5cm]{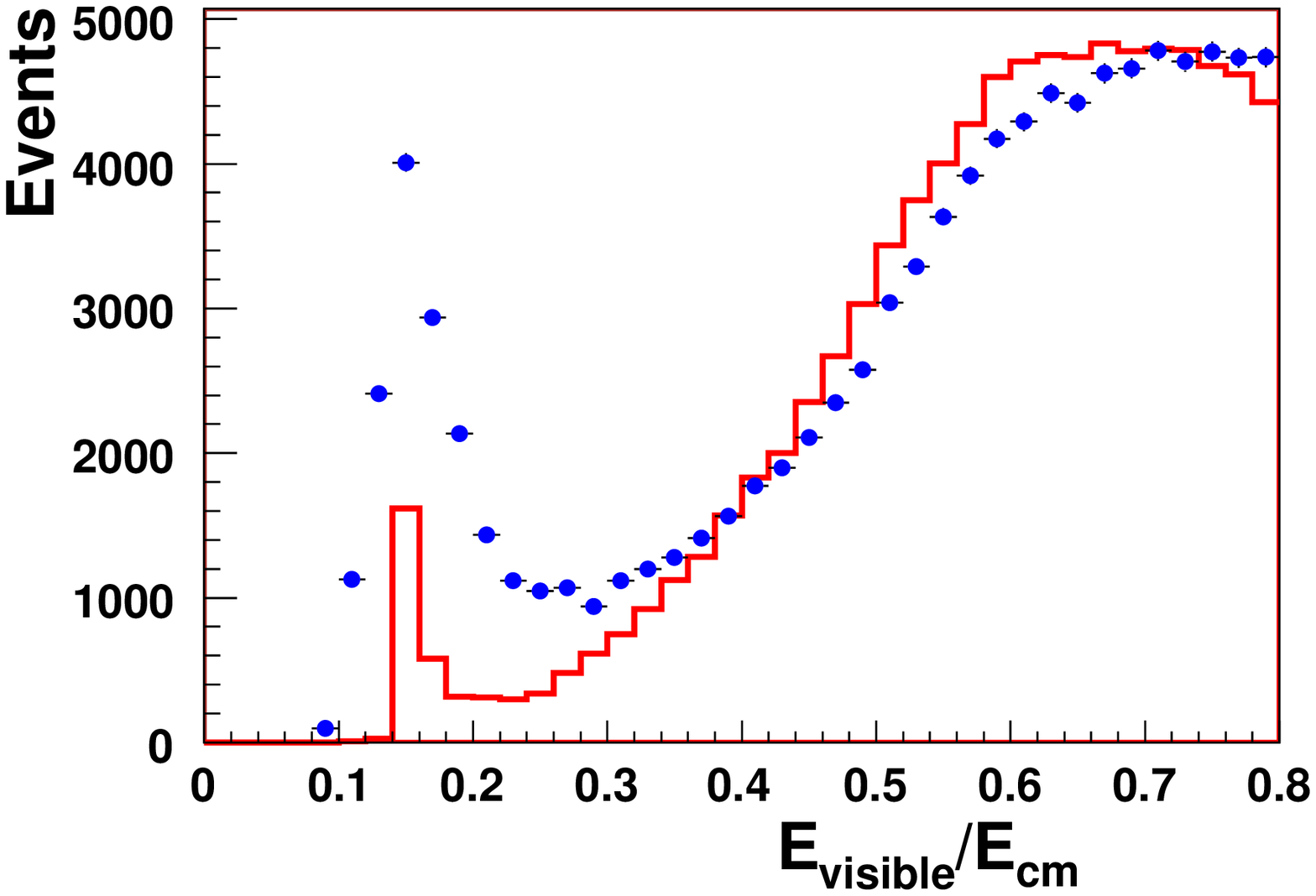}
\caption{\label{evis2prdt}The $E_{visible}/E_{cm}$ distribution
for $N_{good}=2$ events. Dots with error bars are data; the
histogram is MC simulation, normalized to $E_{visible}/E_{cm}>0.4$.}  \efg

\bfg \includegraphics[width=5.0cm,height=4.5cm]{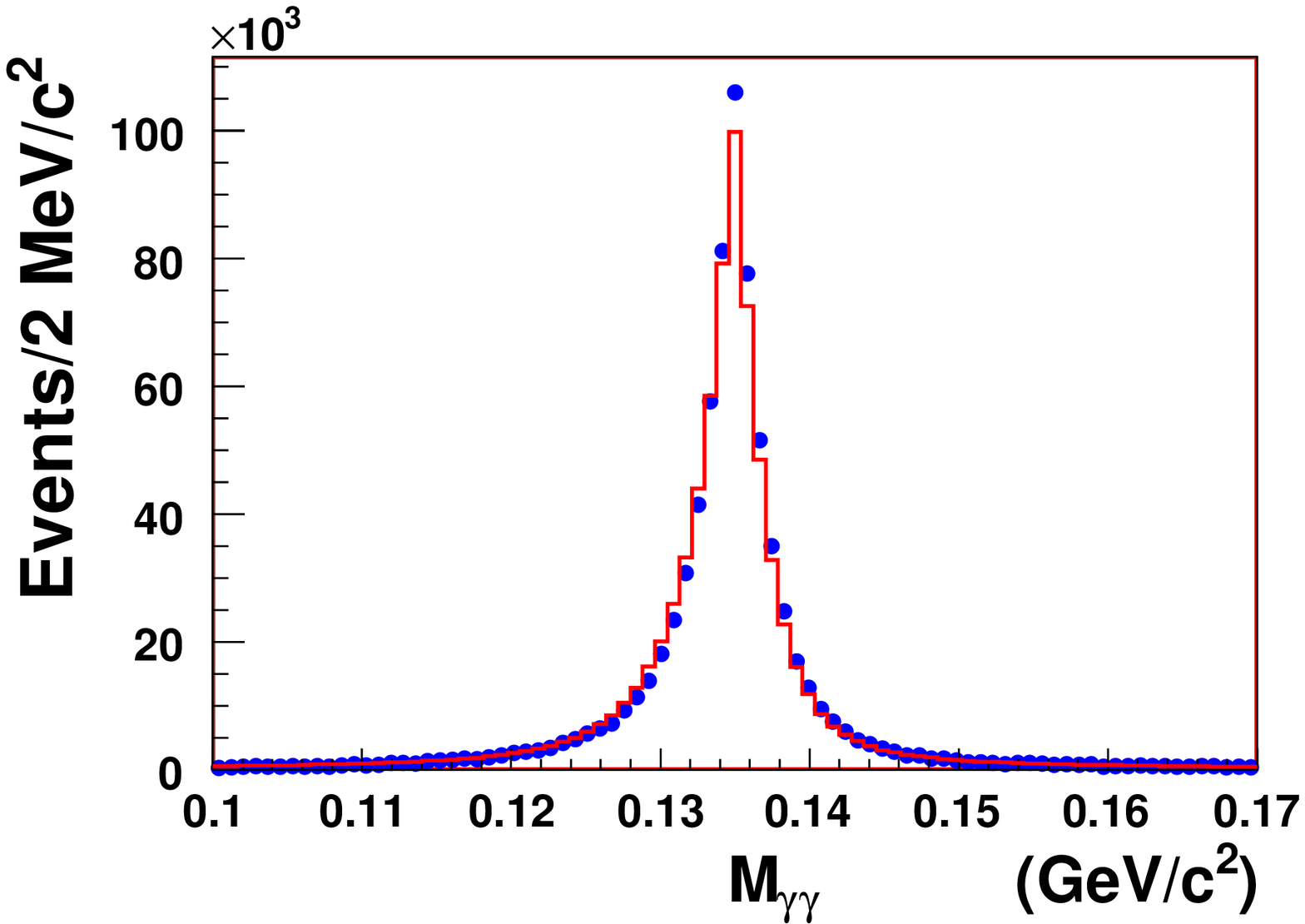}
\caption{\label{pi0}The $\gamma \gamma$ invariant mass
($M_{\GG}$) distribution in the $\pi^0$ mass region for
$N_{good}=1$ events.
 Dots with error bars are data; the histogram is MC simulation.}
\efg

\bfg
\includegraphics[width=5.0cm,height=4.5cm]{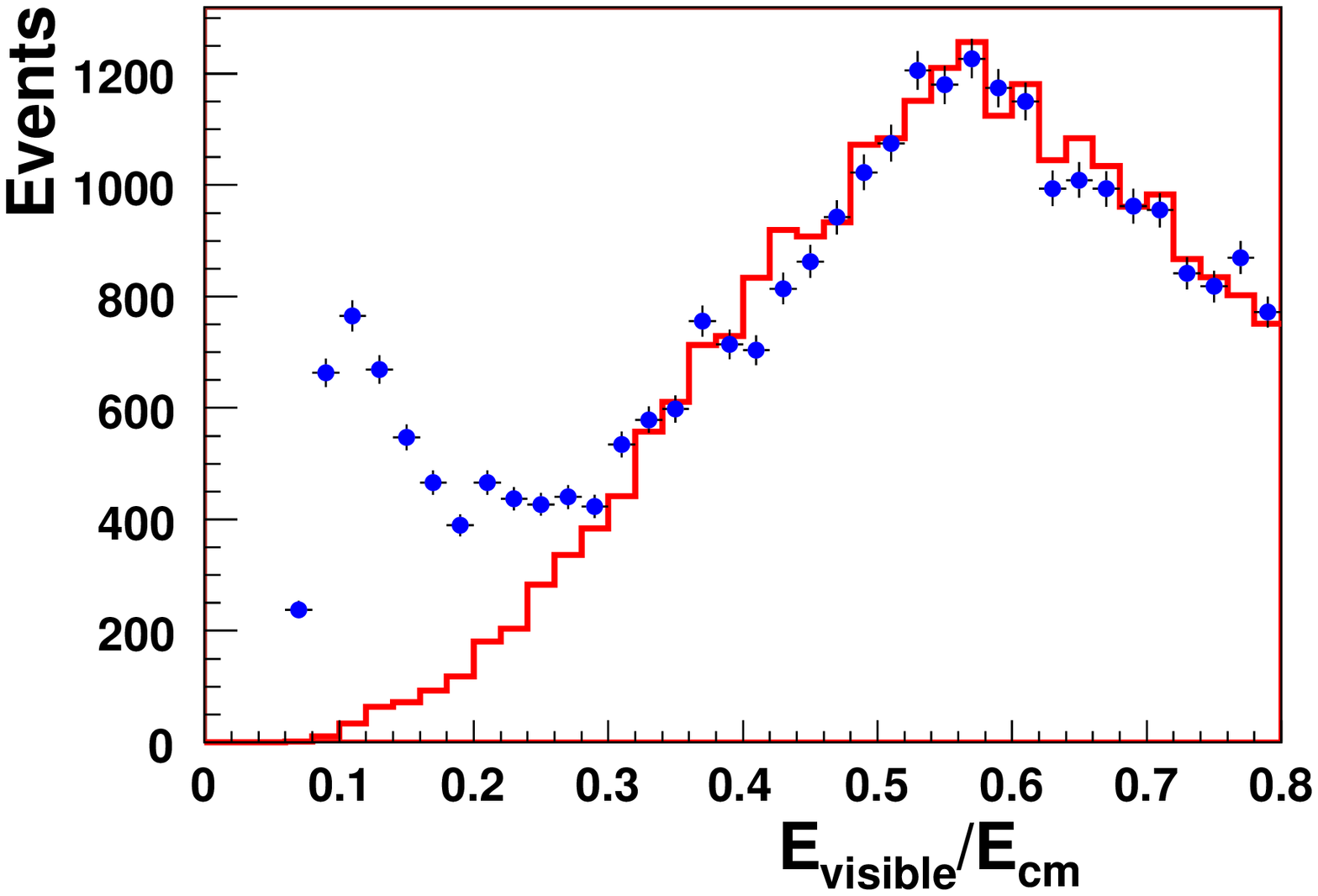}
\caption{\label{evis1prdt}The $E_{visible}/E_{cm}$ distribution
for $N_{good}=1$ events. Dots with error bars are data; the
histogram is MC simulation, normalized to $E_{visible}/E_{cm}>0.3$.}
\efg

The average Z-direction vertex for an event is defined as
\[
\bar{V}_{Z}=\frac{\sum\limits^{N_{good}}_{i=1}V_{Z}^i}{N_{good}},
\]
where $V_Z$ is the distance along the beam direction of the point of
closest approach of a track to the IP.  Figure~\ref{z0dt} shows the
$\bar{V}_{Z}$ distribution for $\psp$ data after the above
selection. Events satisfying $|\bar{V}_Z|<4.0 $ cm are taken as
signal, while events in the sideband region 6.0 cm$<|\bar{V}_Z|<10.0$
cm are taken as non-collision background events. The number of
observed hadronic events ($N^{obs.}$) is determined by
\beq
\label{nhad} N^{obs.}=N_{signal}-N_{sideband}.
\eeq
Another method to determine the number of hadronic events (described below) is
to fit the average Z-vertex with a double Gaussian to describe the
signal and a polynomial to describe the non-collision events.

\bfg \includegraphics[width=5.0cm,height=4.5cm]{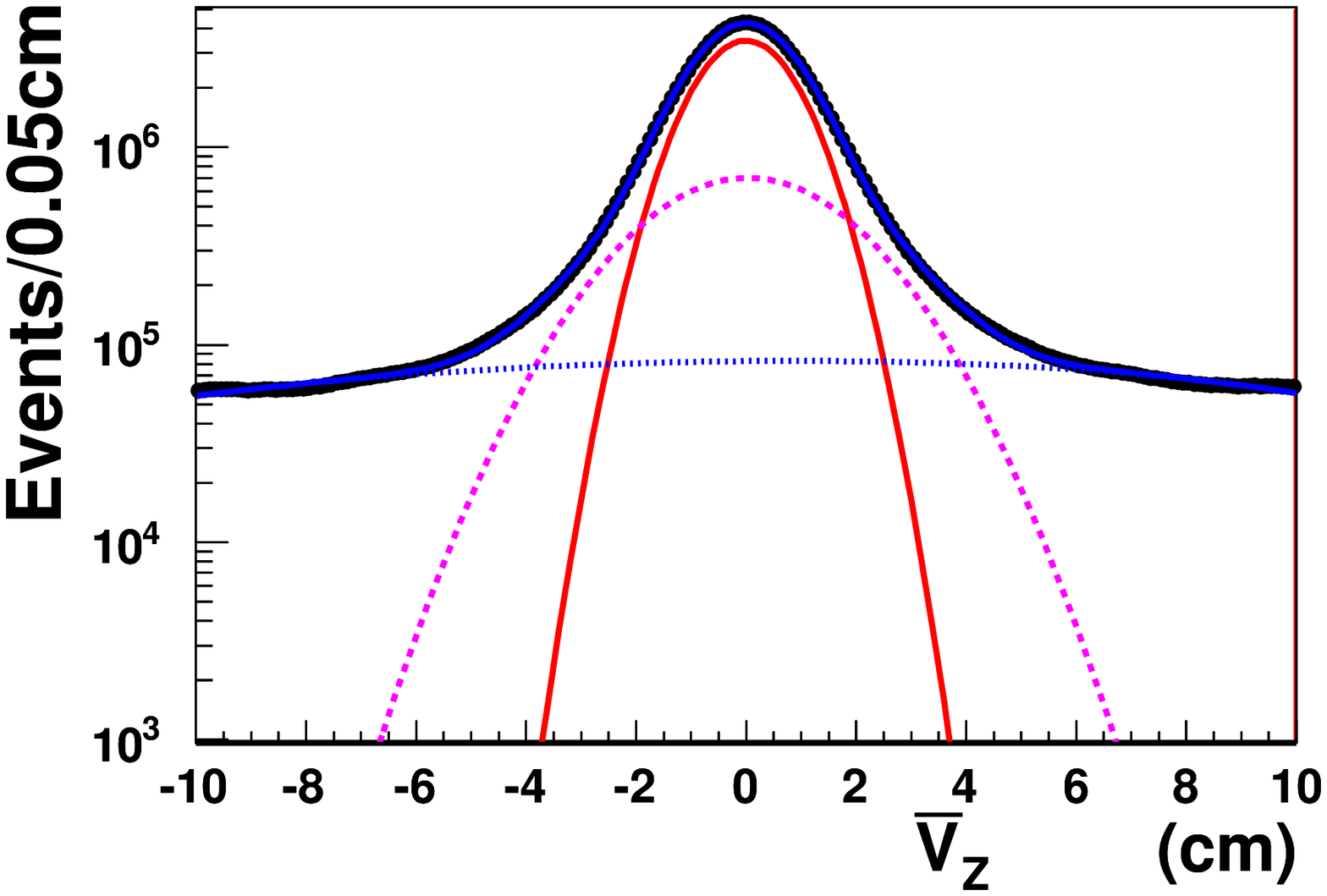}
\caption{\label{z0dt}The average $Z$ vertex ($\bar{V}_{Z}$)
distribution of hadronic events in $\psp$ data.  The curves are a
double Gaussian to describe the signal and a polynomial to
describe the non-collision events.}  \efg

\bfg \includegraphics[width=5.0cm,height=4.5cm]{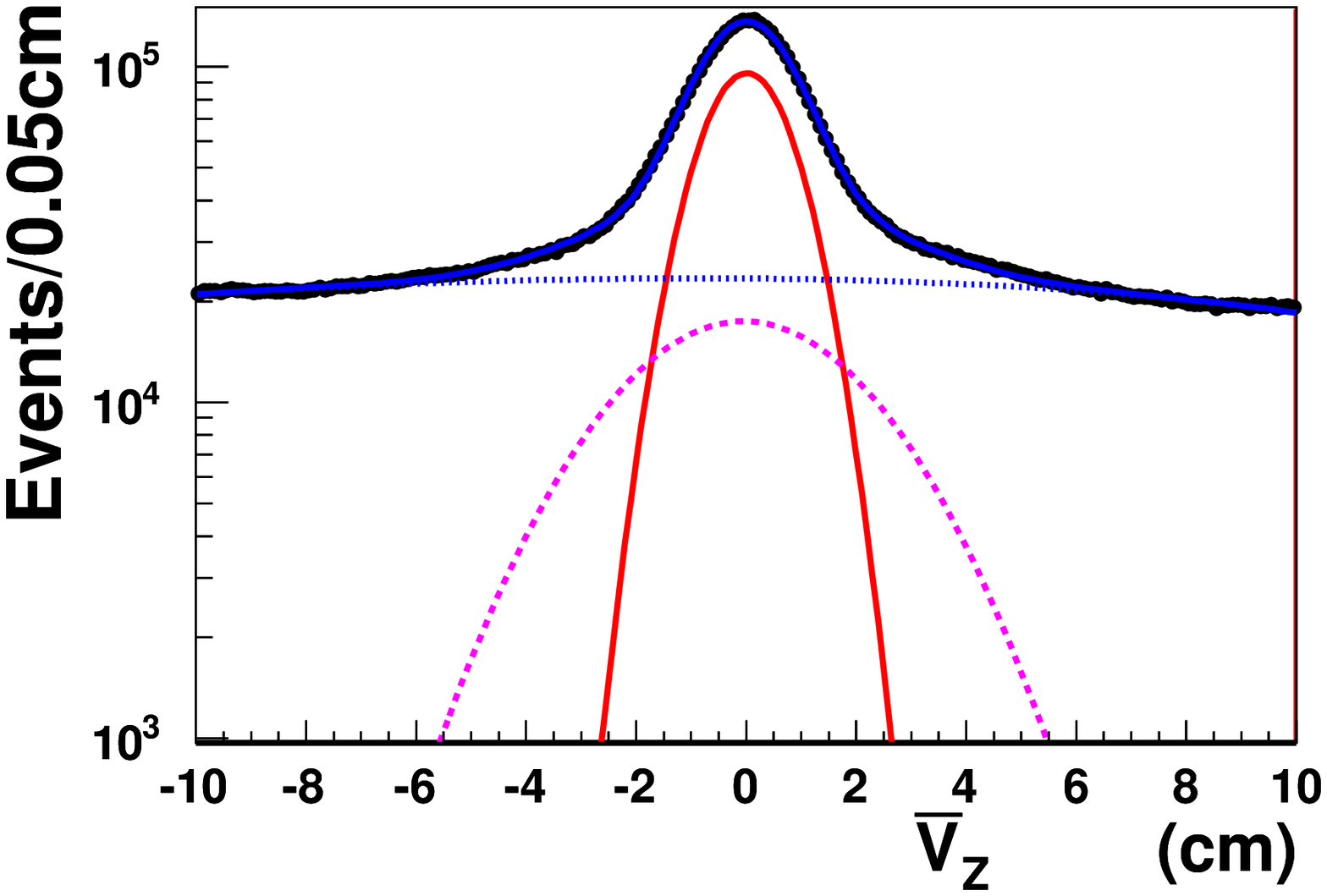}
\caption{\label{z0con}The average $Z$ vertex ($\bar{V}_{Z}$)
distribution of hadronic events in off-resonance data.  The curves
are a double Gaussian to describe the signal and a polynomial to
describe the non-collision events.}  \efg

\section{Background subtraction}

In principle, the number of QED events can be estimated from:
\beq\label{phbg} N^{QED} = {\cal L}\cdot\sigma\cdot\epsilon, \eeq
where ${\cal L}$ is the luminosity, and $\sigma$ and $\epsilon$
are the cross-section and efficiency, respectively.  $\sigma$ is
usually obtained from theoretical prediction, and $\epsilon$ is
determined from MC simulation.

However in this analysis, we use the large sample of off-resonance data collected at 3.65 GeV to estimate the continuum background. The events remaining,
after imposing the same selection criteria in the off-resonance
data, also form a peak in the $\bar{V}_Z$ distribution, as shown
in  Figure~\ref{z0con}. The same signal and sideband regions are
used as for the $\psp$ data to determine the collision and non
collision events.   With this method, the continuum background
subtraction is independent of MC simulation, and little systematic
bias is introduced.

The contributions from radiative returns to $\jpsi$ and $\jpsi$
decays from the extended tail of the Breit-Wigner are very similar
at the $\psp$ peak and off-resonance energy due to the small
energy difference. They are estimated to be 1.11 and 1.03 $nb$ at
the $\psp$ peak and the off-resonance energy point, respectively, and
according to MC simulation, the efficiencies for the known
continuum processes at the two energy points are also similar.
Therefore, the off-resonance data can be employed to subtract both
the continuum QED and $\jpsi$ decay backgrounds using a scaling
factor, $f$, determined from the integrated luminosity multiplied
by a factor of $1/s$ ($s=E_{cm}^2$) to account for the energy
dependence of the cross-section:

\beq\label{factor}
f=\frac{{\cal L}_{\psp}}{{\cal L}_{3.65}}\cdot\frac{3.65^2}{3.686^2}=3.677,
\eeq
where, ${\cal L}_{\psp}$ and ${\cal L}_{3.65}$ are the integrated
luminosities for $\psp$ data and 3.65 GeV data, respectively.

The luminosities at the two different energy points are determined
from $\EE\ar\GG$ events using the same track and event level selection
criteria. At the track level, no good charged tracks and at least two
showers are required. The energy for the most energetic shower should
be higher than $0.7\times E_{beam}$ while the second most energetic
shower should be larger than $0.4\times E_{beam}$, where $E_{beam}$ is
the beam energy. At the event level, the two most
energetic showers in the $\psp$ rest frame should be back to back, and
their phi angles must satisfy $178^{\circ} <|\phi_1 - \phi_2 | <
182.0^{\circ}$. The luminosity systematic errors nearly cancel in
calculating the scaling factor due to small energy difference between
these two energy points.  The $f$ factor can also be obtained using
luminosities determined with Bhabha events. It is found to be 3.685.

Also of concern is the LEB remaining in the $\psp$ events after
the $E_{visible}/E_{cm}$ requirement.  In order to test if the
continuum background subtraction is also valid for these events,
candidate LEB events are selected by requiring
$E_{visible}/E_{cm}<0.35$ where there are few QED events expected.
Figures~\ref{bkg1pr} and ~\ref{bkg2pr} show the comparison of
$E_{visible}/E_{cm}$ between peak and off-resonance data for
$N_{good} =1$ and $N_{good}=2$ events, respectively. The agreement
between the two energy points is good for these events. The ratios
of the numbers of peak and off-resonance events for $N_{good} =1$
and $N_{good}=2$ are 3.3752 and 3.652, respectively. Compared with
the scaling factor obtained from luminosity normalization in
Eq.(~\ref{factor}), a difference of about 10\% is found for
$N_{good}=1$ while there is almost no difference for $N_{good}=2$
events. These differences will be taken as systematic errors.

The small numbers of events from $\psp\ar\EE,~\MM$, and $\TT$ in data
that pass our selection do not need to be explicitly subtracted since
$\psp \to lepton$ events are included in the inclusive MC and those
passing the selection criteria will contribute to the MC determined
efficiency, so that their contribution cancels.

\bfg \includegraphics[width=5.0cm,height=4.5cm]{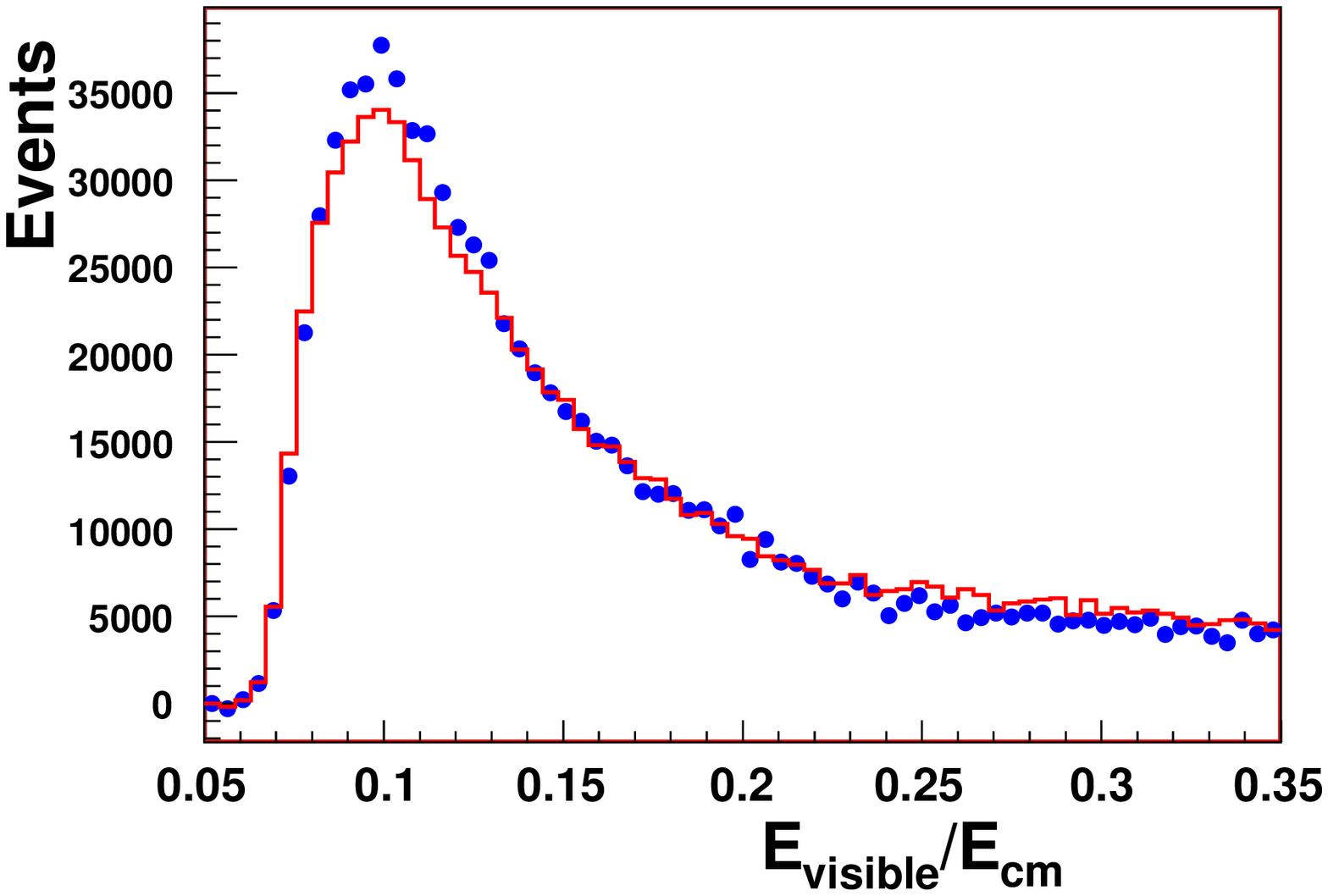}
\caption{\label{bkg1pr}Comparison of LEB events between $\psp$
peak and off-resonance data for $N_{good}=1$ events. Dots with
error bars are $\psp$ data, and the histogram is off-resonance
data.} \efg

\bfg \includegraphics[width=5.0cm,height=4.5cm]{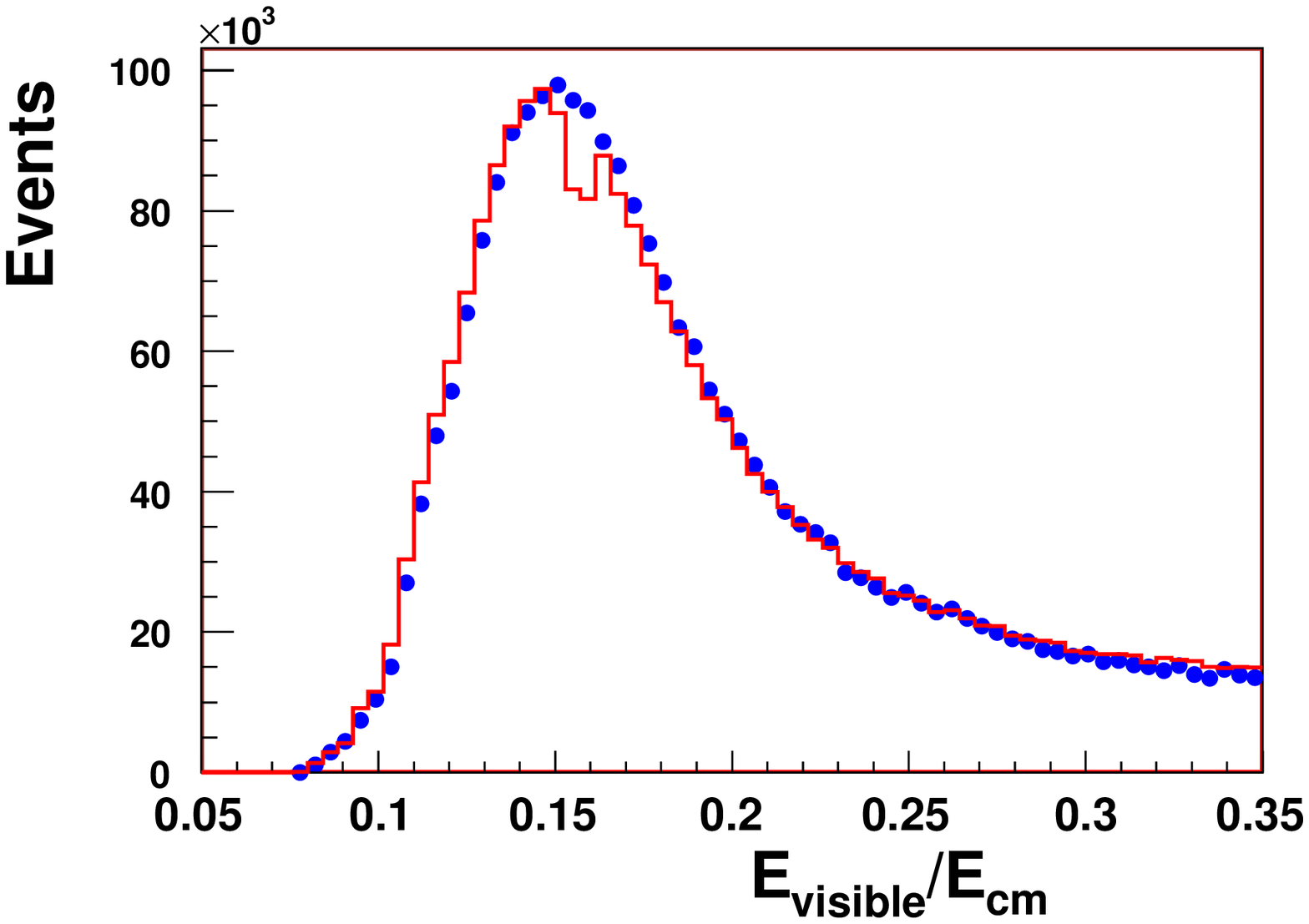}
\caption{\label{bkg2pr}Comparison of LEB events between $\psp$
peak and off-resonance for $N_{good}=2$ events. Dots with error
bars denote $\psp$ data, and the histogram denotes off-resonance
data.} \efg

Table~\ref{npsppart} shows the number of observed hadronic events
for different multiplicity requirements for $\psp$ and
off-resonance data. Figures~\ref{cost}, ~\ref{etot}, and
~\ref{ncharge} show the $\cos\theta$, $E_{visible}$, and
charged-track multiplicity distributions after subtracting
background.

\btbl
\caption{$N^{obs.}$ for peak and off-resonance data ($\times 10^6$), and the detection
efficiency for inclusive $\psp$ decay events determined with
$106\times 10^6$ $\psp\ar~inclusive$ MC events.}
\bcl
\doublerulesep 2pt
\begin{tabular}{lcccc}\\\hline\hline
&$N_{good}\geq 1$&$N_{good}\geq 2$&$N_{good}\geq 3$&$N_{good}\geq 4$\\\hline\hline
$\psp$ data&106.928&102.791&81.158&63.063\\
off-resonance data&2.192&1.98&0.704&0.433\\
$\epsilon$(\%)&92.912&89.860&74.624&58.188\\\hline
\end{tabular}
\label{npsppart}
\ecl
\etbl

\bfg
\includegraphics[width=5.0cm,height=4.5cm]{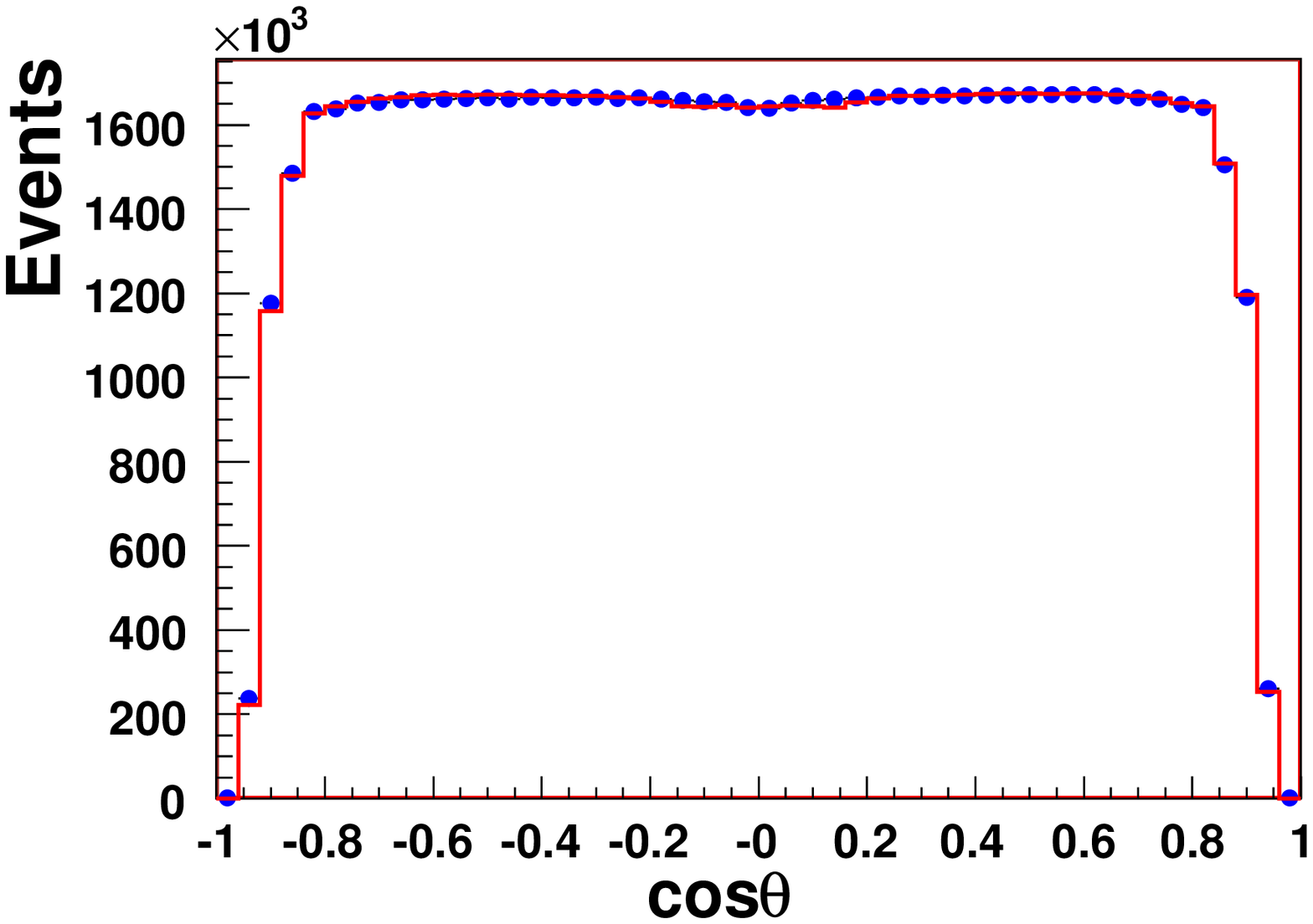}
\caption{\label{cost} The $\cos\theta$ distribution for charged
tracks. Dots with error bars are data; the histogram is MC simulation.}
\efg

\bfg
\includegraphics[width=5.0cm,height=4.5cm]{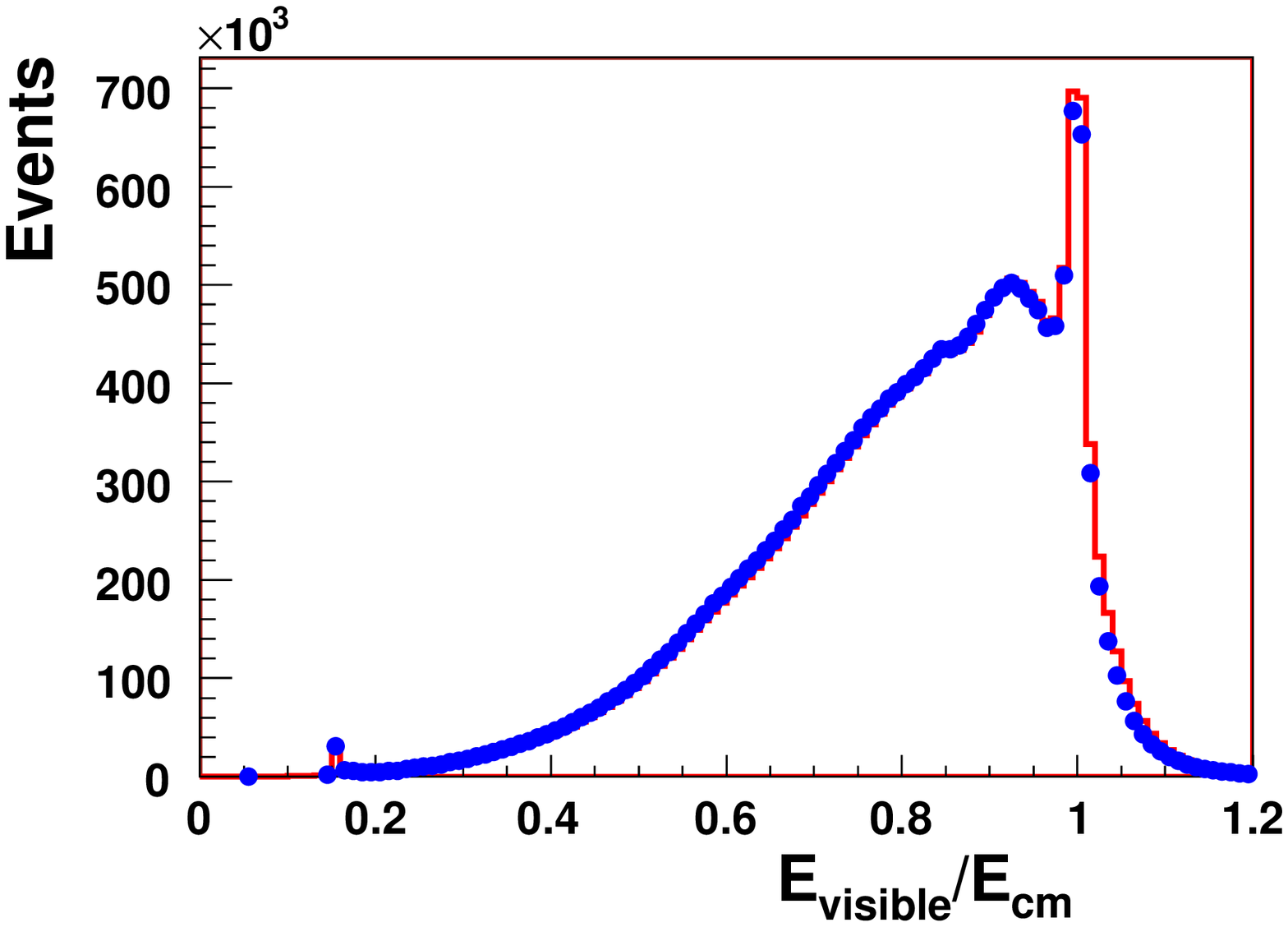}
\caption{\label{etot}The visible energy distribution.  Dots with
error bars are data; the histogram is MC simulation.}  \efg

\bfg
\includegraphics[width=5.0cm,height=4.5cm]{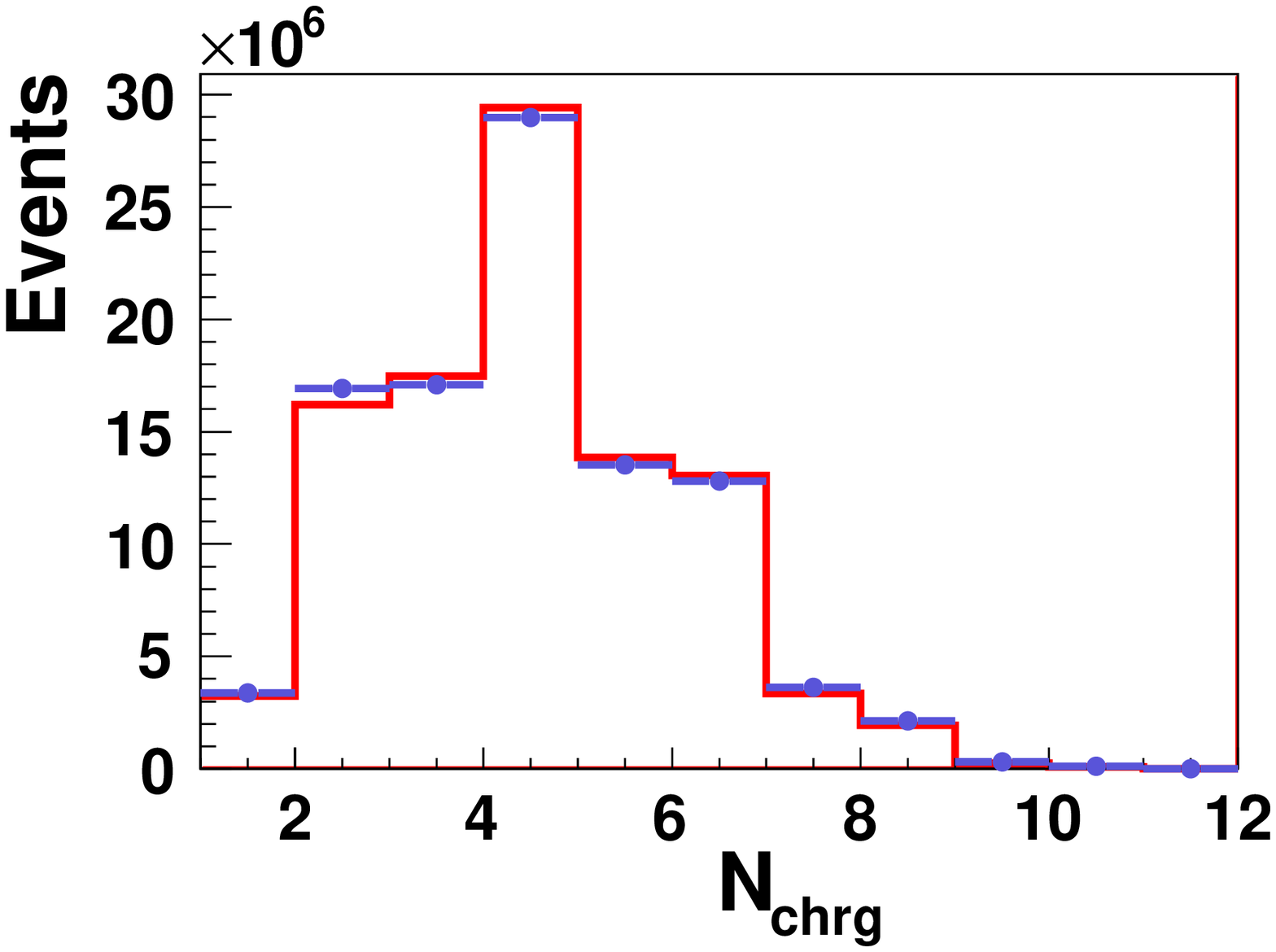}
\caption{The charged-track multiplicity distribution. Dots with
error bars are data; the histogram is MC simulation.} \label{ncharge} \efg

\section{Numerical result}
The number of $\psp$ events is determined from \beq\label{npsip}
N_{\psp}=\frac{N_{peak}^{obs.}-f\cdot
N_{off-resonance}^{obs.}}{\epsilon}, \eeq where, $N_{peak}^{obs.}$ is
the number of hadronic events observed at the $\psp$ peak from
Eq.~(\ref{nhad}), $N_{off-resonance}^{obs.}$ is the number of hadronic
events observed at the off-resonance energy point, $E_{cm}=3.650$ GeV,
with the same selection criteria as those for peak data, and $\epsilon$
is the selection efficiency obtained from the inclusive $\psp$
MC sample, the branching fraction of $\psp\ar~inclusive~hadron$ is included in the efficiency. The relevant numbers are listed in Table~\ref{npsppart} for
different $N_{good}$ selection requirements.  The factor $f$ is the
scaling factor which has been introduced in Eq.~(2). With these
numbers, we obtain the numerical result for $\npsp$ listed in
Table~\ref{totno} for different choices of $N_{good}$. We take the
result for $N_{good}\geq 1$ as the central value of our final result.

\btbl
\caption{$\npsp$ ($\times 10^6$) for different
charged-track multiplicity requirements.}
\bcl
\doublerulesep 2pt
\begin{tabular}{ccccc}\hline\hline
&$N_{good}\geq 1$&$N_{good}\geq 2$&$N_{good}\geq 3$&$N_{good}\geq 4$\\\hline
$N_{\psp}$~~&106.414&106.279&105.289&105.643\\\hline
\end{tabular}
\label{totno}
\ecl
\etbl

\section{Systematic Uncertainties}
The systematic uncertainties include the uncertainties caused by
tracking, the event start time ($T_0$), trigger efficiency, background
contamination, the selection of the signal and sideband regions, etc.
\subsection{Tracking}
Generally, the tracking efficiency for MC events is higher than
that of data according to various studies~\cite{track}. Assuming
the average efficiency difference between data and MC is 1\% per
track, the effect can be measured by randomly tossing out 1\% of
MC simulated tracks. Only a difference of 0.03\% on $\npsp$ is
found for $N_{good}\geq 1$ events with and without this tracking
efficiency change; $\npsp$ is not sensitive to the tracking
efficiency.

\subsection{Charged-track multiplicity}
Figure~\ref{ncharge} shows that the MC does not simulate the
charged-track multiplicity very well. The error due to charged-track
multiplicity simulation can be estimated by an unfolding method, which
is described as follows. The generated true charged multiplicity in MC
simulation is even, i.e., 0,~2,~4,~6,~8, $\cdots$.  The observed MC
multiplicity distribution is obtained after simulation and event
selection. For example, if the generated true multiplicity is 4, the
observed multiplicities are 0,~1,~2,~3, or 4 with different
probabilities. Therefore, an efficiency matrix, $\epsilon_{ij}$, which
describes the efficiency of an event generated with $j$ charged tracks
to be reconstructed with $i$ charged tracks, is obtained from MC simulation.  The
distribution of the number of observed charged-track events in data,
$N_i^{obs.}$, is known. The true multiplicity distribution in data can
be estimated from the observed multiplicity distribution in data and
the efficiency matrix by minimizing the $\chi^2$. The $\chi^2$ is
defined as \beq \chi^2 =
\sum\limits^{10}_{i=1}\frac{(N_i^{obs.}-\sum\limits_{j=0}^{10}\epsilon_{ij}\cdot
N_j)^2}{N_i^{obs.}}, \eeq where the $N_j~(j=0,~2,~4,~6,~8,~10)$ describe the true
multiplicity distribution in data and are taken as floating
parameters in the fit. The simulation is only done up to a true
multiplicity of 10, since there are few events at high
multiplicity. The total true number of events in data can be obtained
by summing all fitted $N_{j}$; it is $105.96\times 10^6$ which is
lower than the nominal value by 0.4\%. We take this difference as the
error due to the charged-track multiplicity distribution.

\subsection{Momentum and opening angle}
For $N_{good}=2$ events, momentum and opening angle requirements are
used to remove the huge number of Bhabha events.  When the momentum
requirement is changed from $P<1.7$ GeV/$c$ to $P<1.55$ GeV/$c$, the
corresponding $N^{obs.}$ for peak and resonance data, as well as the
efficiency change, but the change in $\npsp$ is only 0.05\%. When the
angle requirement is changed from $\theta<176^{\circ}$ to $\theta<160^{\circ}$,
the change in $\npsp$ is 0.01\%.  Therefore, the total uncertainty due
to momentum and opening angle requirements is 0.05\%.
Figures~\ref{pt2prh} and ~\ref{angep2pr} show comparisons between data
and MC simulations for momentum and opening angle distributions after
background subtraction, respectively.

\bfg \includegraphics[width=5.0cm,height=4.5cm]{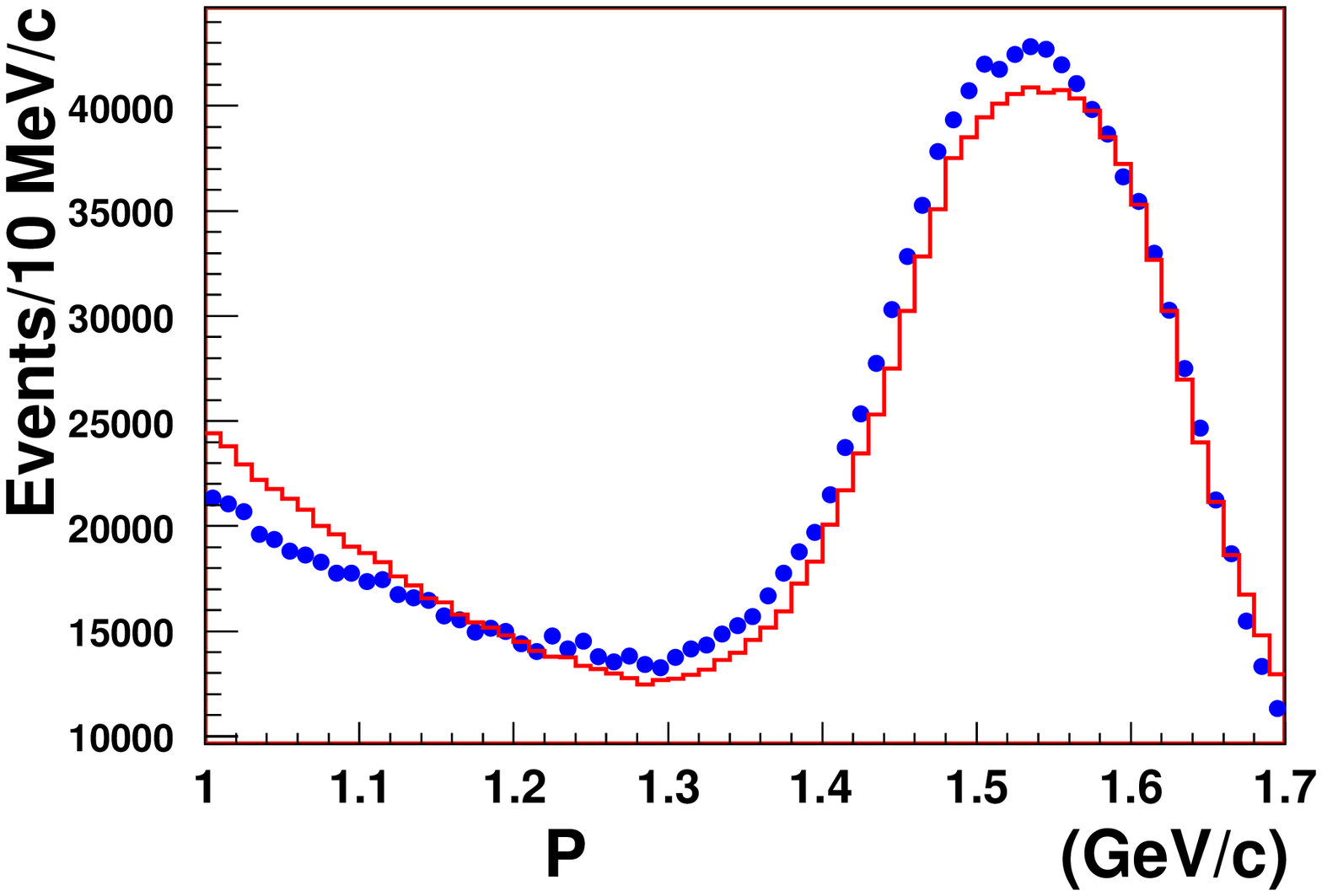}
\caption{\label{pt2prh} The distribution of total momentum for
$N_{good}=2$ events. Dots with error bars are data; the histogram
is MC simulation.}
\efg
\bfg
\includegraphics[width=5.0cm,height=4.5cm]{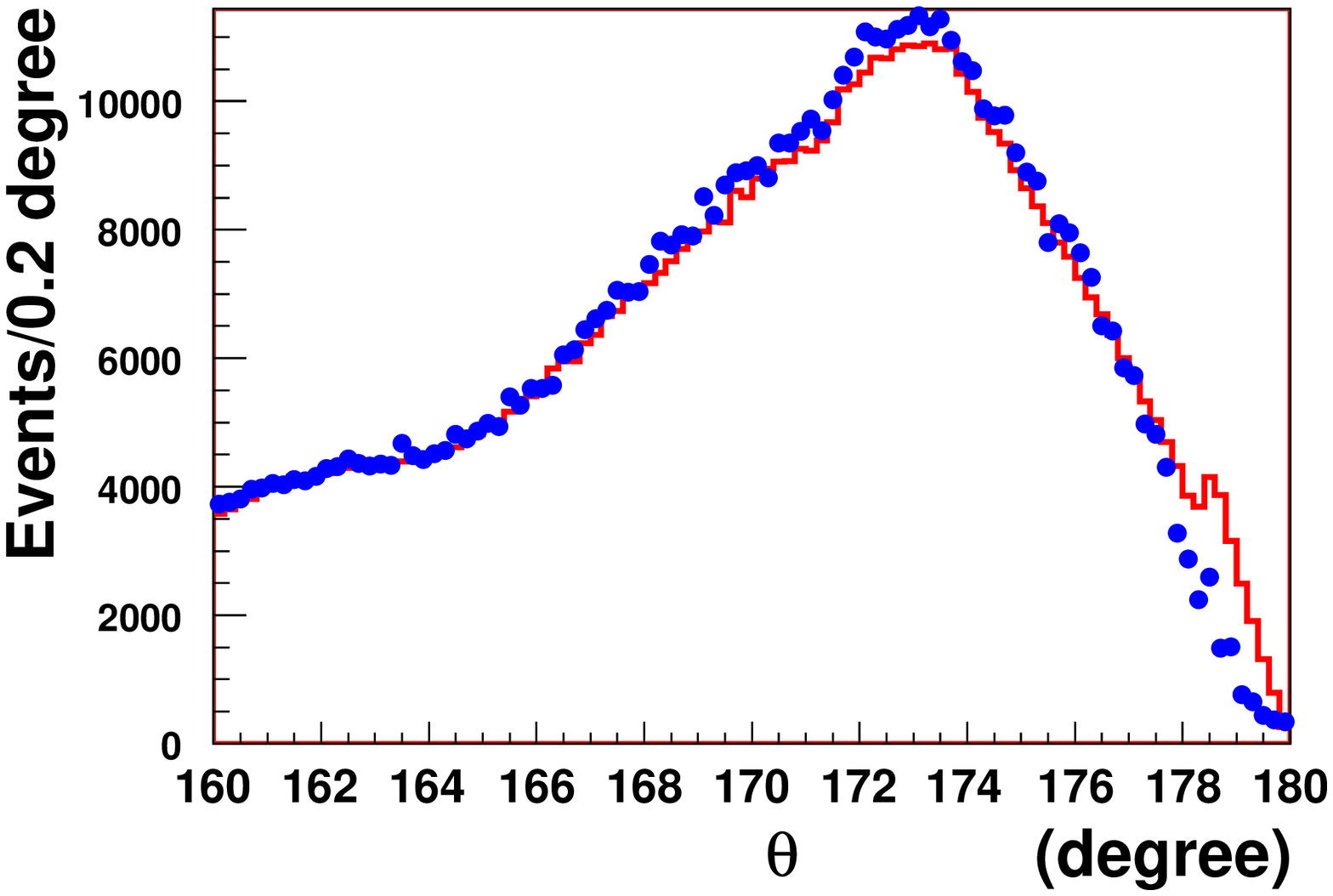}
\caption{\label{angep2pr} The distribution of opening angle
between tracks for $N_{good}=2$ events. Dots with error bars are
data; the histogram is MC simulation.}  \efg

\subsection{LEB background contamination}
$\npsp$ is insensitive to the visible energy requirement. The
difference between a tight requirement, $E_{visible}/E_{cm}>0.45$, and
no requirement is only 0.1\%. Conservatively, an error of 0.1\% is
assigned due to the background contamination.

\subsection{Determination of number of hadronic events}
Two methods are used to obtain $N^{obs.}$. The first is to directly
count the numbers of events in the signal and sideband regions; the
second method is to fit the $\bar{V}_Z$ distribution with a double
Gaussian for the signal and a polynomial for the background, as shown
in Figs.~\ref{z0dt} and \ref{z0con}. A difference of 0.28\% is found
between these two methods which is taken as the error due to the
uncertainty from the $N^{obs.}$ determination.

\subsection{Vertex limit}
When $V_r<1$ cm is changed to $V_r<2$ cm, $\npsp$
changes by 0.35\%, while if $|\bar{V}_Z|<10$ cm is changed to
$|\bar{V}_Z|<15$ cm, there is almost no change. Therefore, the
difference of 0.35\% is taken as the error from the vertex
requirement.

\subsection{Scaling factor}
The scaling factor can be obtained for two different QED processes,
$\EE\ar\GG$ and $\EE\ar\EE$. The corresponding results are 3.677 and
3.685. The difference on $\npsp$ due to the $f$ factor can be calculated
by $\Delta f\cdot N^{obs.}_{N_{good}\geq 1}(3.650$~GeV$)/N_{\psp}
=(3.685-3.677)\cdot 3.1808/106.32 = 0.023\%$. The slight difference
indicates the uncertainty caused by the normalization factor is
negligibly small.

\subsection {Choice of sideband region}
We take $|\bar{V}_Z|<4.0$ cm as the signal region and
$6<|\bar{V}_Z|<10$ cm as the sideband region. A difference of 0.45\%
in $N_{\psp}$ is found by shifting the sideband region outward by 1.0
cm, which is about 1$\sigma$ of the $\bar{V}_Z$ resolution, i.e., the
sideband region is changed from 6 cm$ <|\bar{V}_Z|<10$ cm to 7cm
$<|\bar{V}_Z|<11$ cm.  We take this difference as the error due to the
uncertainty caused by choice of the sideband region.

\subsection {\boldmath $\pi^0$ mass requirement}
This requirement is only used for $N_{good}=1$ events.  $\npsp$ has a
slight change of 0.11\% when the mass window requirement is
changed from $|M_{\GG}-M_{\pi^0}|<0.015$ GeV/$c^2$ to
$|M_{\GG}-M_{\pi^0}|<0.025$ GeV/$c^2$; this difference is taken as the
uncertainty due to $\pi^0$ mass requirement.

\subsection {\boldmath The cross section of $\EE\ar\TT$}
Since the off-resonance energy point is not very far from $\tau \tau$
threshold, $\sigma({\EE\ar\TT})$ does not vary as $1/s$ between the
off-resonance energy and the $\psp$ peak, as other QED processes. The
difference between the observed and the cross section assuming a $1/s$
dependence causes a change of 0.17\% in $\npsp$. This change is taken
as a systematic error.

\subsection{\boldmath $B(\psp\ar X+\jpsi)$}
The $\psp$ MC assumes $B(\psp\ar X+\jpsi)\approx 57\%$ from the
PDG~\cite{PDG}, while the CLEO experiment determined a branching
ratio of 62\%~\cite{cleo}. Using CLEO's result, a new inclusive MC
sample was generated. The corresponding efficiencies are 92.912\%,
89.761\%, 74.838\% and 58.528\% for $N_{good}\geq 1,~ 2, ~3$ and
4, respectively. Compared with numbers in Table~\ref{npsppart},
the efficiency differences between these two MC samples are
negligible.

\subsection{Event start time determination}

The Event Start Time (EST) algorithm is used to determine the common
start time of the recorded tracks in an event. The efficiency of the
EST determination affects the resolution of tracks from the tracking
algorithm. These efficiencies for different charged tracks, $e$,
$\mu$, $\pi$, $K$, and $p$, and photons are studied with different
control samples for both data and inclusive MC events, for example,
$\jpsi\ar\ppp$, $\pp\ppb$, and $\psipto, \jpsi\to
l^+l^-$, etc.  All comparisons indicate that the efficiencies of the
EST determination are high for both track and event level $(>98\%)$
selection, and the difference between data and MC simulated events is quite
small $(\sim 0.1\%)$. We take this difference as the uncertainty
caused by the EST determination.

\subsection{Trigger efficiency}

The fraction of events with $N_{ngood}\geq 2$ is about 97\%. The
trigger efficiency for these events is close to 100.0\% according
to a study of the trigger efficiency~\cite{nik}. For $N_{good}=1$
events, an extra $\pi^0$ is required, and the hadron trigger
efficiency for this channel is 98.7\%~\cite{nik}. Since the
fraction of $N_{good}=1$ events is only about 3\%, the uncertainty
caused by the trigger is negligible.

\subsection{The missing 0-prong hadronic events}
A detailed topology analysis is performed for $N_{good}=0$ events
in the inclusive MC sample. Most of these events come from known
decay channels, such as $\psp\ar X+\jpsi~(X=
\eta,~\pi^0\pi^0,~$and$ ~\pp)$, $\psp\ar\gchicJ$, and
$\psp\ar\EE$, $\MM$. The fraction of pure neutral events is less
than 1.0\%. For the known charged decay modes, the MC simulation
works well according to many comparisons between data and MC
simulation in Section 3. To investigate the pure neutral channels,
the same selection criteria at the track level are used. The
criteria at the event level include $N_{good}=0$ and
$N_{\gamma}>3$. The latter requirement is used to suppress
$\EE\ar\GG$ and beam-associated background events. The same
selection criteria are imposed on the off-resonance data.
Figures~\ref{0prdt} and ~\ref{0prmc} show the distribution of
total energy in the EMC for data and inclusive MC events. The
peaking events correspond to the pure neutral candidates, and the
number of events is extracted by fitting. The difference in the
number of fitted events between data and inclusive MC events is
found to be 17\%. Therefore, the uncertainty due to the pure
neutral events should be less than $17\%\times 1\%= 0.17\%$, and
this is taken as the systematic error on the missing 0-prong
events.

\bfg
\includegraphics[width=5.0cm,height=4.5cm]{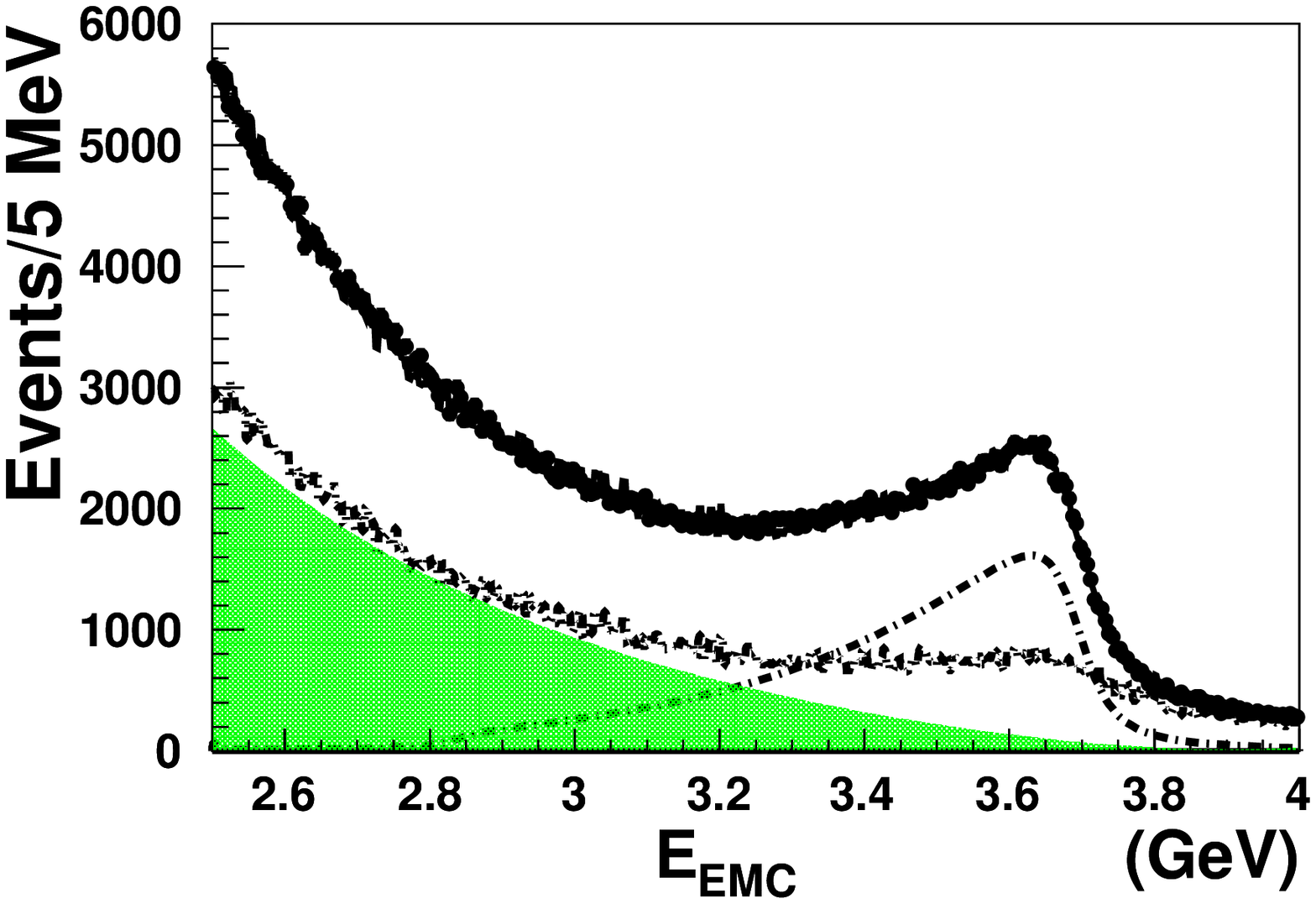}
\caption{\label{0prdt}The distribution of total energy in the EMC with
$N_{good}=0$ for data. The dot-dashed line denotes the signal shape of $\psp\ar~neutral~channel$, the dashed line denotes the background shape from QED processes, and the shaded region is the background shape from $\psp$ decay.}  \efg

\bfg
\includegraphics[width=5.0cm,height=4.5cm]{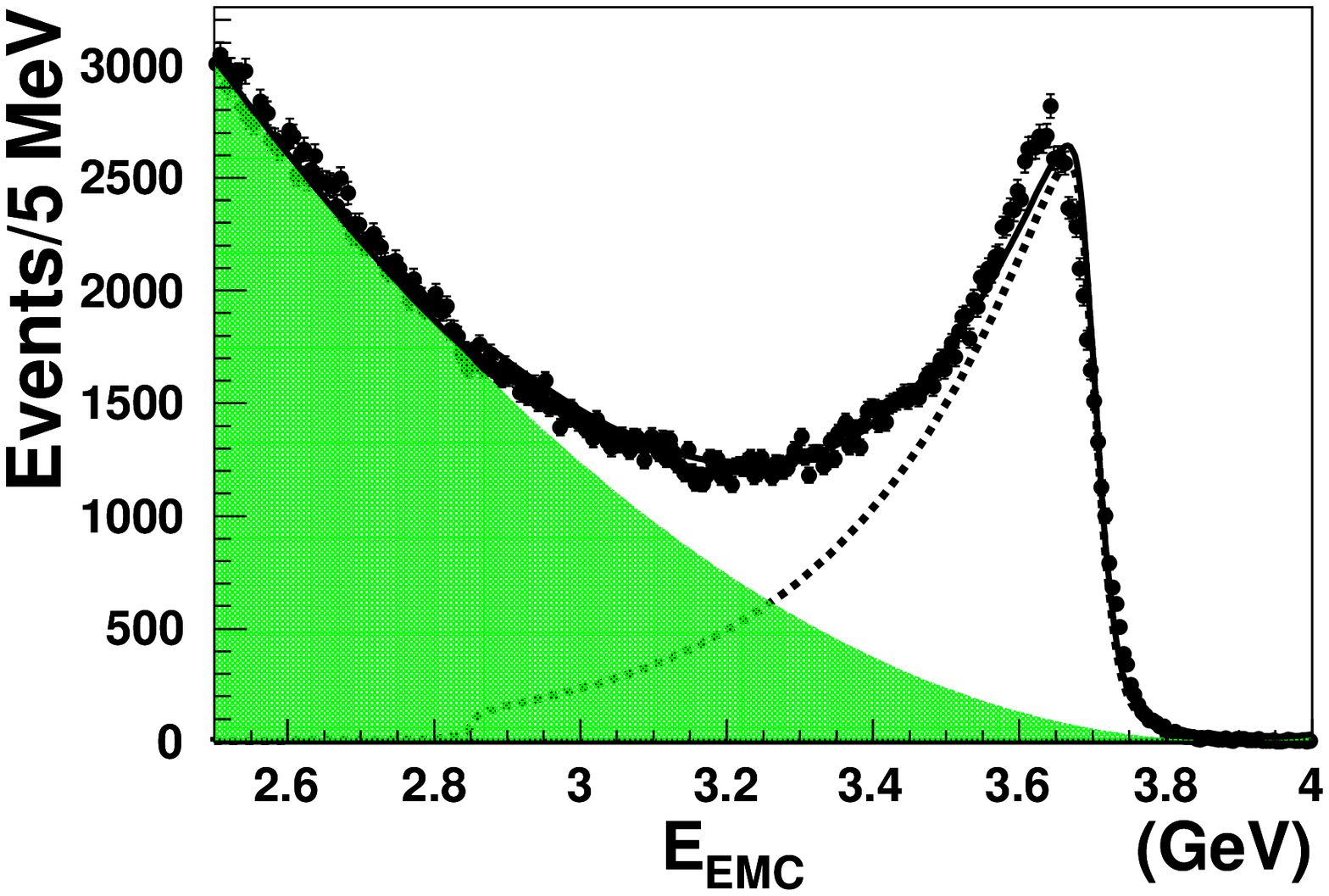}
\caption{\label{0prmc}The distribution of total energy in the EMC with $N_{good}=0$ for inclusive MC events. The dashed line denotes the signal shape of $\psp\ar~neutral~channel$, and the shaded region is the background shape from $\psp$ decay.}
\efg

\subsection{\boldmath $B(\psp\ar hadrons)$}
The uncertainty of $B(\psp\ar hadrons)$ is very small
according to the PDG~\cite{PDG}, 0.13\%, which is taken as the error due to
uncertainty of $\psp$ decays to hadronic events.

\subsection{Total error}
Table~\ref{bkg} lists all systematic errors. The total systematic
error is determined by the quadratic sum of all errors.

\btbl
\caption{The systematic error (\%)}
\bcl
\doublerulesep 2pt
\begin{tabular}{ccccc}\hline\hline
 Source&Error\\\hline\hline
Background contamination&0.10\\
$N^{obs.}$ determination&0.28\\
Choice of sideband region&0.45\\
Vertex selection &0.35\\
Momentum and opening angle&0.05\\
Scaling factor ($f$) & 0.02\\
0-prong~ events&0.17\\
Tracking & 0.03\\
Charged-track multiplicity&0.40\\
$\sigma(\EE\ar\TT)$&0.17\\
$B(\psp\ar X+\jpsi)$&0.00\\
$\pi^0$ mass requirement &0.11\\
EST determination&0.10\\
Trigger efficiency&negligible\\
$B(\psp\ar hadron)$&0.13\\\hline\hline Total&0.81\\\hline\hline
\end{tabular}
\label{bkg}
\ecl
\etbl

\section{Summary}
The number of $\psp$ events is determined using $\psp\ar
~hadrons$. The large off-resonance data sample at $E_{cm}=3.65$
GeV is used to estimate the background under the $\psp$ peak. The
number of $\psp$ events taken in 2009 is measured to be
$(106.41\pm0.86)\times 10^6$, where the error is systematic only
and the statistical error is negligible.

\section{Acknowledgment}
The BESIII collaboration thanks the staff of BEPCII and the
computing center for their hard efforts. This work is supported in
part by the Ministry of Science and Technology of China under
Contract No. 2009CB825200; National Natural Science Foundation of
China (NSFC) under Contracts Nos. 10625524, 10821063, 10825524,
10835001, 10935007, 11125525, 10975143, 10979058; Joint Funds of the National
Natural Science Foundation of China under Contracts Nos. 11079008,
11179007; the Chinese Academy of Sciences (CAS) Large-Scale
Scientific Facility Program; CAS under Contracts Nos.
KJCX2-YW-N29, KJCX2-YW-N45; 100 Talents Program of CAS; Istituto
Nazionale di Fisica Nucleare, Italy; Ministry of Development of
Turkey under Contract No. DPT2006K-120470; U. S. Department of
Energy under Contracts Nos. DE-FG02-04ER41291, DE-FG02-91ER40682,
DE-FG02-94ER40823; U.S. National Science Foundation; University of
Groningen (RuG) and the Helmholtzzentrum fuer Schwerionenforschung
GmbH (GSI), Darmstadt; WCU Program of National Research Foundation
of Korea under Contract No. R32-2008-000-10155-0.

\end{document}

%% file: def-com.tex
\newcommand{\chicJ}{\chi_{cJ}}
\newcommand{\gchicJ}{\gamma\chi_{cJ}}

\newcommand{\psipto}{\psi^{\prime}\rightarrow \pi^+\pi^- J/\psi}

\newcommand{\ppp}{\pi^+\pi^- \pi^0}

\newcommand{\EE}{e^+e^-}
\newcommand{\MM}{\mu^+\mu^-}
\newcommand{\GG}{\gamma\gamma}
\newcommand{\TT}{\tau^+\tau^-}
\newcommand{\pp}{\pi^+\pi^-}

\newcommand{\ppb}{p\bar{p}}

\newcommand{\psp}{\psi^{\prime}}
\newcommand{\jpsi}{J/\psi}
\newcommand{\ar}{\rightarrow}

\newcommand{\npsp}{N_{\psp}}

\newcommand{\bfg}{\begin{figure}}
\newcommand{\efg}{\end{figure}}
\newcommand{\bitm}{\begin{itemize}}
\newcommand{\eitm}{\end{itemize}}
\newcommand{\bnum}{\begin{enumerate}}
\newcommand{\enum}{\end{enumerate}}
\newcommand{\btbl}{\begin{table}}
\newcommand{\etbl}{\end{table}}
\newcommand{\btbu}{\begin{tabular}}
\newcommand{\etbu}{\end{tabular}}
\newcommand{\bcl}{\begin{center}}
\newcommand{\ecl}{\end{center}}
\newcommand{\bbt}{\bibitem}
\newcommand{\beq}{\begin{equation}}
\newcommand{\eeq}{\end{equation}}
\newcommand{\beqr}{\begin{eqnarray}}
\newcommand{\eeqr}{\end{eqnarray}}